\documentclass[12pt]{article}

\usepackage{graphicx}
\usepackage{times,fullpage}
\usepackage{amsmath,amssymb}
\usepackage{color,setspace}

\title{Professional diversity and the productivity of cities} 

 \author{Lu\'is M. A. Bettencourt,$^{1}$ Horacio Samaniego,$^{2,3}$\footnote{To whom correspondence should be addressed; E-mail:  horacio@ecoinformatica.cl}   HyeJin Youn,$^{1}$\\
 \\
 \normalsize{$^{1}$  Santa Fe Institute, 1399 Hyde Park Rd, Santa Fe NM 87501, USA,}\\
 \normalsize{$^{2}$  Center for Non Linear Studies, Theoretical Division MS B284, }\\  
 \normalsize{  Los Alamos National Laboratory, NM 87545, USA,}\\
 \normalsize{$^{3}$Facultad de Ciencias Forestales \& Recursos Naturales, }\\
 \normalsize{Universidad Austral de Chile, Valdivia, Chile}
 \\
 \normalsize{}
 }

\date{}

\begin{document}

\maketitle

\begin{abstract}
  The relationships between diversity, productivity and scale determine much of the structure and robustness of complex biological and social systems\cite{May2001,Bonner1988}.
  While arguments for the link between specialization and productivity are common\cite{Smith2010,Henderson1974,Glaeser1992,Henderson1995,PugaGillesDuranton2000}, diversity has often been invoked as a hedging strategy\cite{Haldane-May,Scheffer2012}, allowing systems to evolve in response to environmental change\cite{Haldane-May,Scheffer2012,Quigley1998}. Despite their general appeal, these arguments have not typically produced quantitative predictions for optimal levels of functional diversity consistent with observations.
 One important reason why these relationships have resisted formalization is the idiosyncratic nature of diversity measures, which depend on given classification schemes\cite{May1988, Marshall2010}.   
  Here, we address these issues by analyzing the statistics of professions in cities and show how their probability distribution takes a universal scale-invariant form, common to all cities, obtained in the limit of infinite resolution of given taxonomies. 
We propose a model that generates the form and parameters of this distribution via the introduction of new occupations at a rate leading to individual specialization subject to the preservation of access to overall function via their ego social networks. 
  This perspective unifies ideas about the importance of network structure in ecology and of innovation as a recombinatory
  process with economic concepts of productivity gains obtained through the division and coordination of labor, stimulated by scale.
\end{abstract}

\doublespacing

\newpage

A fundamental theme across many complex systems\cite{May2001,Bonner1988} - from ecosystems\cite{Tilman2001} to human behavior\cite{Eagle2010} and socio-economic organization\cite{Hidalgo2007b, Hidalgo2009} - deals with understanding the mechanisms by which diversity arises and is sustained. In contemporary human societies socioeconomic diversity is associated primarily with cities\cite{Jacobs1961} accounting for their role in producing new ideas and stimulating development\cite{Jacobs1961, Quigley1998, PugaGillesDuranton2000}.  However, counter arguments have also been made noting that specialized cities are sometimes more productive\cite{Henderson1974, Glaeser1992, Henderson1995, PugaGillesDuranton2000}. Familiar examples are contemporary Silicon Valley or manufacturing cities in their heyday.  Nevertheless, these questions remain far from settled, in part because of the difficulties inherent to measuring diversity in any complex system\cite{May1988, Marshall2010}.

Measures of diversity typically account for the presence, and sometimes the relative proportion\cite{Hirschman1945,Cover1991,Mori2008}, of different functional types, for example different professions or business types in cities or nations, or different species in an ecosystem\cite{Magurran2003}. Such measures, are inevitably linked to particular classification schemes or taxonomies.  To appreciate this point consider the question: How many different professions are there in a large city, like New York?  In general, there is no objective answer to this question as it depends on how finely one differentiates similar functions.  Here, we show that under specific conditions a limit of infinite resolution can be obtained in a way similar to the treatment of physical quantities close to phase transitions\cite{Stanley1987} and that, in this limit, scheme-independent measures of diversity can emerge. 

The simplest measure of diversity, $D(N)$, counts the number of distinct professions present in a city.  Fig. 1A shows $D$, for US metropolitan areas vs. their total employment, $N_e$. Because $N_e$ is, on average, proportional to population\cite{Bettencourt2007}, $N$, we use the two measures of scale interchangeably. $D$ increases with $N_e$ initially and then saturates for large cities and is well fit by
\begin{equation}
  D(N_e) = d_0 \frac{\left(\frac{N_e}{N_0}\right)^\gamma} {1+\left(\frac{N_e}{N_0}\right)^\gamma}.
  \label{eq:1}
\end{equation}
Eq.~1 holds over time and for different levels of resolution, $r$, in the occupations hierarchical classification scheme  (see SI).  The parameters in Eq.~1 are, in general, functions of $r$. The scale $d_0(r)$ is the effective size of the classification scheme at resolution $r$, $N_0(r)$ is a characteristic size of a city at which saturation starts. $\gamma$, empirically independent of $r$ (see SI), is a scaling exponent giving the proportionality between the population growth rate and that of new occupations in the city, in the absence of saturation.  

Coarsening the hierarchical classification leads to similar saturation at each of the scheme's size $d_0(r_6), d_0(r_5), d_0(r_4)$, etc (Fig. 1B).  This behavior is the hallmark of a finite resolution artefact, a phenomenon well understood in terms of finite size scaling at phase transitions\cite{Stanley1987}. 
The explicit dependence of $d_0$ on $r$ means that given classification schemes are too coarse to capture the professional diversity of large US cities\cite{Mameli2008}, beyond $N_0\sim 10^5$. Nevertheless, we can use the variation of the statistics of occupations with $r$ to derive classification scheme independent results.   We reconcile all curves for $D(N)$ at different $r$ and extract their limit as $r \rightarrow \infty$. We define a dimensionless function $h(N$ $/$ $N_0,\gamma)$ such that
\begin{equation*}
  D(N)= d_0 ~h\left(\frac{N}{N_0}\right) ~\left( \frac{N}{N_0} \right)^\gamma  \rightarrow \quad
  \begin{cases}
   D_0~ N^\gamma, \qquad N << N_0,  \\
    d_0(r),\qquad   \quad N >> N_0, 
     \end{cases}
\label{eq:2}
\end{equation*}
where $D_0$ is a constant.  Comparison with Eq.~1 tells us that in the limit $\frac{N}{N_0} \rightarrow 0$, $h \rightarrow 1$, $D_0 \rightarrow \frac{d_0}{N_0^\gamma}$, and in the limit $\frac{N}{N_0} \rightarrow \infty$, $h\rightarrow \left( \frac{N_0}{N} \right)^\gamma$. A universal scaling regime exists if and only if the quantity $D_0=\frac{d_0(r)}{N_0(r)^\gamma}$ becomes a constant, independent of $r$, as $r\rightarrow \infty$ (Fig. 1B). Fig. 1C shows $d_0$ \textit{vs.}  $N_0^\gamma$ across $r$ and over time.  The relationship is well described by a straight-line with slope of $D_0=0.05$ across all years.  These results suggest the existence of a resolution independent, scale-invariant limit for $D(N)$ and show that the occupational diversity of cities is in fact open-ended: the number of distinct occupations in US cities increases by $\sim85\%$ with each doubling of its labor force, meaning that larger cities are at once more diverse in absolute terms and more specialized per capita.  These insights can be proven as simple theorem (see SI).

Beyond analyzing the presence or absence of professions, which gives only a crude measure of urban diversity, we can characterize their frequency distribution.  The analysis of the frequency of different types in complex systems, from word frequency to city size, is naturally described in terms of their (Zipfian) rank-frequency distribution.  To derive this distribution we identify $D(N)$ with the maximum rank at each value of $N$, which has probability $p(D) = \frac{1}{N}$.  Inverting this relation and generalizing it to all ranks, $i$, leads to the occupational frequency, $f(i)$:
\begin{eqnarray}
  f(i)   = \frac{N_e}{N_0} \left( \frac{d_0 - i}{i} \right)^{1/\gamma}.
\label{eq:prob_dist}
\end{eqnarray}
This is also scheme independent in the large resolution limit and can be used to derive the probability density, $p(i)$, as
\begin{eqnarray}
  p(i) = \frac{f(i)}{\sum_{i=1}^D f(i) }= \frac{1-\gamma}{\gamma} \frac{i^{-1/\gamma}}{1 - D(N)^{-\frac{1-\gamma}{\gamma}}}; 
\label{eq:prob_occ}
\end{eqnarray}
which is also independent of $r$. The occupational probability has a residual dependence on $N$ through $D(N)$ because the rarest professions cannot have less than one person. This is the only source of city size dependence of traditional measures of diversity such as the Herfindahl-Hirschman index or the Shannon entropy (see SI)\cite{Hirschman1945,Cover1991,Magurran2003}, which are functionals of $p(i)$.  Given Eq.~3, both measures express increases in diversity (the Herfindahl-Hirschman index decreases, the Shannon entropy increases) towards a finite limit at infinite $N$. For large cities the approach to this limit is controlled by a term $\sim N^{\delta}$, with $\delta=1-\gamma$ (see SI for derivation).

Fig. 2 shows that the distribution of occupations for different cities is universal: When adjusted for scale, $N_e$, all frequency curves collapse onto a single line.  This shows that there is an expected nested sequence of occupations, predicted by city size, as expected by the hierarchy principle of central place theory\cite{Christaller1966,Mori2008} and in analogy to products {\it vs.} level of economic development at the national level\cite{Hidalgo2007b,Hidalgo2009}.    

A simple model that predicts the form of the occupational diversity distribution, Eq.~3, is a version of the Yule-Simon mechanism of preferential attachment\cite{Simon1955,Zanette2005}:  as the city grows by one more job, $\Delta N_e=1$, it creates a  new occupation with probability $\alpha= \frac{d D} {d N_e}=\gamma D_0 N_e^{\gamma-1}$, or it takes up an existing profession, proportionally to its frequency, with probability  $1-\alpha$. For large $N_e$ this predicts an exponent\cite{Simon1955,Zanette2005} in the occupational distribution of $ \gamma=\frac{N_e}{1-\alpha(N_e)} \frac{ \alpha(N_e)}{D(N_e)} $, note that even for small $N_e$, $\alpha< 0.04 <<1$.  

Given the results so far, we may expect economic productivity to be inversely proportional to professional diversity. Consider that indicators of economic productivity (wages, GDP) scale, on average, superlinearly\cite{Bettencourt2007} with $N$, $W(N,t) = W_0(t) N(t)^\beta$, with $W_0(t)$ and $\beta \simeq 1+\delta>1$ independent of $N$ (see SI). 
An average wage per capita is, then,  $w(N)=W_0 N^\delta$, where $\delta\sim 1/6\simeq 1-\gamma$.  This result has been derived from a general theoretical framework that defines cities as co-located social networks, subject to infrastructural efficiency constraints\cite{Bettencourt2012}, with $w=G k(N)$, where $G$ is a constant in $N$, involving a balance between people and infrastructural properties, and $k(N)=k_0 N^\delta$ is the average social connectivity (degree) per person, which has been observed in urban telecommunication networks\cite{Schlapfer2012}. Similarly, diversity per capita, $d(N) =D(N)/N = D_0 N^{\gamma-1}= D_0 N^{-\delta}$.  Hence, we conclude that $w \sim 1/d$.   This relation is an expression of the abundant evidence in economics for specialization (a decrease in $d(N)$)  as the source of increases in productivity\cite{Smith2010,Henderson1995}. However, no city has become rich by reducing its occupational diversity to a single activity:  What then is the optimal level of diversity that maximizes the economic productivity of a city?

To answer this question we observe that the process of specialization, by which an individual sheds tasks to others, requires that such functions remain tightly integrated so that overall functionality is preserved.   This implements a form of comparative advantage at the individual level, where increases in productivity at each node, gained through specialization, remain integrated with other necessary functions via social network links.  We may therefore require that the number of functions directly accessible to each person is preserved as the city grows. The number of functions that each individual reaches directly through its social network is $N_f = d . k$, which we require to stay constant, $A$, in $N$, $d. k = A$.   We now reconcile the expectation that $w\sim 1/d$ with the claim that, like other urban socioeconomic outputs, $w$ be proportional to social connectivity\cite{Bettencourt2012}, $w\sim k$. We write $w = g(k d )/d$, with $g$ an analytic function, independent of $N$. We now maximize wages subject to the conservation of functionality across social links by a Lagrange multiplier procedure (see SI) to show that $d = A/k$ and that $w(N) = g(A)/d(N) = g(A)/A ~k(N)$.  Then,  $D(N) = A/k(N) N = A/k_0 N^{1-\delta}=D_0 N^\gamma$, which predicts the form of the scaling of occupational diversity with city size. As shown above, this relation, taken across all $N$, also predicts the rank-size distribution of urban occupations and associated measures of diversity.  

In summary, we showed that the patterns of occupational diversity and economic productivity observed in US metropolitan areas can be derived from an integrated view of cities as socioeconomic networks that promote a systematic division and coordination of labor without loss of overall functionalities available to individuals.  Similar quantitative patterns characterize the technological complexity of simpler human societies\cite{Kline2010} and may be a property of networked systems that can experience open-ended increases in their productivity with scale. The reversibility of these processes, e.g. the existence of hysteresis in the externalization and reabsorption of functions by individuals and networks, may also underlie the resilience of many complex systems\cite{Tilman2001,Marshall2010,Scheffer2012} under unexpected functional change or population loss.

\subsubsection*{Acknowledgements}
  We thank Doug Erwin, Ricardo Hausmann, Cesar Hidalgo, Jos\'e Lobo and Geoffrey West for discussions. This research
  is partially supported by the Rockefeller Foundation, the James S. McDonnell Foundation (grant no. 220020195), the
  National Science Foundation (grant no. 103522), the John Templeton Foundation (grant no. 15705), the U.S. Department
  of Energy through the LANL/LDRD Program (contract no. DE-AC52-06NA25396), and by a gift from the Bryan J. and June
  B. Zwan Foundation.

\bibliographystyle{naturemag}

% % \bigskip
% % \noindent
% % The authors declare that they have no competing financial interests.

% \bigskip

% \noindent
% Correspondence and requests for materials should be addressed to
% H.S.~(email: horacio@ecoinformatica.cl).

\clearpage
\begin{center}
  \includegraphics[width=17.8cm]{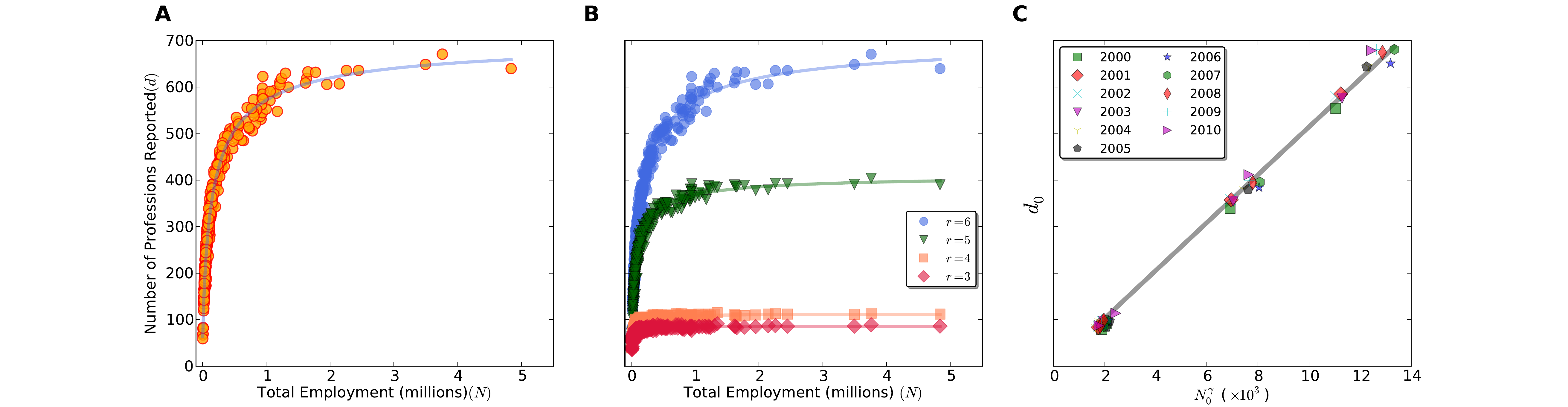}
\end{center}
\paragraph*{Fig. 1.} The number of distinct occupations in US Metropolitan Statistical Areas {\it vs.} total
employment. \textbf{(A)} The relationship between the number of professions present for each city (orange dots) and city
size is well described by $D(N_e)=d_0 \frac{(N_e/N_0)^\gamma}{1+(N_e/N_0)^\gamma}$, with $d_0=686$, $\gamma=0.84$,
$N_0=1.48\times 10^5$ (blue line). \textbf{(B)} $D(N_e)$ at different labels of resolution of the occupational
classification scheme, $r_i$, with $i=6$ the finest and $i=3$ the coarsest. \textbf{(C)} $d_0$ is proportional to
$N_0^\gamma$ across levels of classification scheme resolution and time, suggesting the there is a $r$-independent limit
to the form of the occupational diversity of cities and that $D$ is open-ended. In this limit, $D(N_e) = D_0 N_e^\gamma$
and larger cities are always more diverse as a whole, but more specialized per capita.

\begin{center}
  \includegraphics[width=8.7cm]{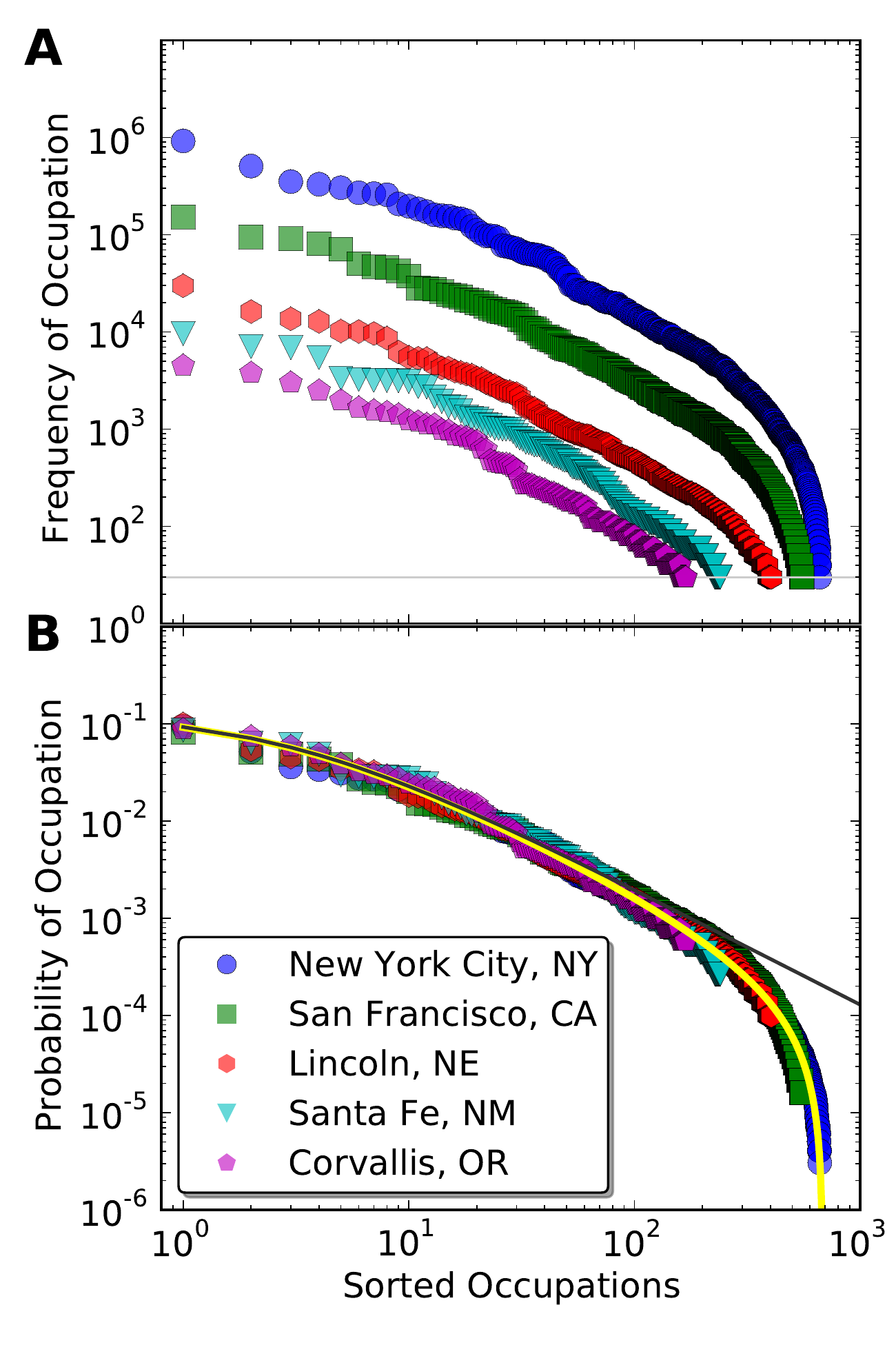}
\end{center}
\paragraph*{Fig. 2.} The distribution of occupations in US metropolitan areas is universal. \textbf{(A)} Frequency
distribution for several cities with different population sizes only differ in their amplitude, which is set by city
size and the extent to which they probe rare occupations. The horizontal grey line shows the minimum number of
professions (thirty) reported.  \textbf{(B)} The rank-probability distributions for different cities collapse on each
other when adjusted for city size (total employment). The yellow line shows the fit of the universal form to
$f(i)/N_e=\frac{1}{N_0} \left(\frac{d_0 - (i+i_0)}{(i+i_0)}\right)^{1/\gamma}$, where we introduces a scale $i_0\simeq 3$
at small ranks. The black line is the form of $f(i)/N_e$ in the absence of saturation.

\end{document}

\maketitle 
\baselineskip24pt

\section*{Supporting Information Text}

\paragraph*{Data} 
We adopt a definition of functional cities in terms of metropolitan statistical areas (MSAs).  MSAs are collections of
political units (counties) aggregated by the US census bureau based on a set of criteria that includes population size,
density and commuting flows. There were about 403 MSAs in 2010, providing an ample basis for studying occupational
diversity across city size (with 50K to 20M inhabitants) as well as any other urban characteristics. MSAs are integrated
labor markets and the best official definition of functional cities in terms of a mixing population
\cite{Bettencourt2012}.

Data on professional occupations in US MSAs was obtained from the Bureau of Labor Statistics (BLS) \cite{OES2011} 
and is
freely available online (http://www.bls.gov/oes/). All occupations in the U.S. economy are hierarchically classified on
the basis of their similarities at different levels of aggregation based on the BLS's Standard Occupational
Classification (SOC) scheme, which contains a total of 811 distinct professions at its finest ($r=r_6$ or 6-digit) level of resolution
for 2010.

\paragraph*{Fit methodology}
For each year, we estimated parameters of Eq.~1 at resolution $r_i$ for $i=$6, 5, 4, and 3 digits level occupations
(Fig. 1A-B) through ordinary least squares \cite{R_Dev_Team2011} , using the Gauss-Newton method, which relies on linear
approximations to the nonlinear mean function \cite{Bates1988}. We then used $\gamma, d_0, N_0$ parameters estimated for
2010 to fit Eq.~2. A small constant $i_0=3$ is introduced to Eq.~2 to account for the initial curvature observed for
most common occupations (Fig.~2B). Hence the fit in Fig.~2B) corresponds to
\begin{equation} 
f(i) = \frac{N_e} {N_0} \left[ \frac{d_0 - (i+i_0)}{i+i_0} \right]^{1/\gamma}.
\end{equation}
Note that $i_0$ is not determined by the process of analytic continuation, valid only at high ranks, that we used to obtain the form of the frequency distribution and constitutes from that point of view a functional freedom that is motivated also by the Yule-Simon model \cite{Simon1955}.

\subsection*{Asymptotic relation between diversity and city size holds over time}

The relation between diversity of employment and the size of urban area is maintained across time. The diversity of professions in the urban area is well described by the relation:
\begin{equation}
  \label{eq:div}
    d(N) = d_0 \frac { \left ( \frac{N}{N_0} \right)^\gamma } {1+ \left ( \frac{N}{N_0} \right)^\gamma },
\end{equation}
where $d_0$ is the effective size of the classification scheme (maximum number of occupations), $N_0$ is the size of the city where saturation starts to occur and $\gamma$ the characteristic exponent describing how $d$ increases with city size, in the absence of saturation.

\bigskip

\begin{figure}[H]
  \centering
  \mbox{
    \includegraphics[width=0.33\textwidth]{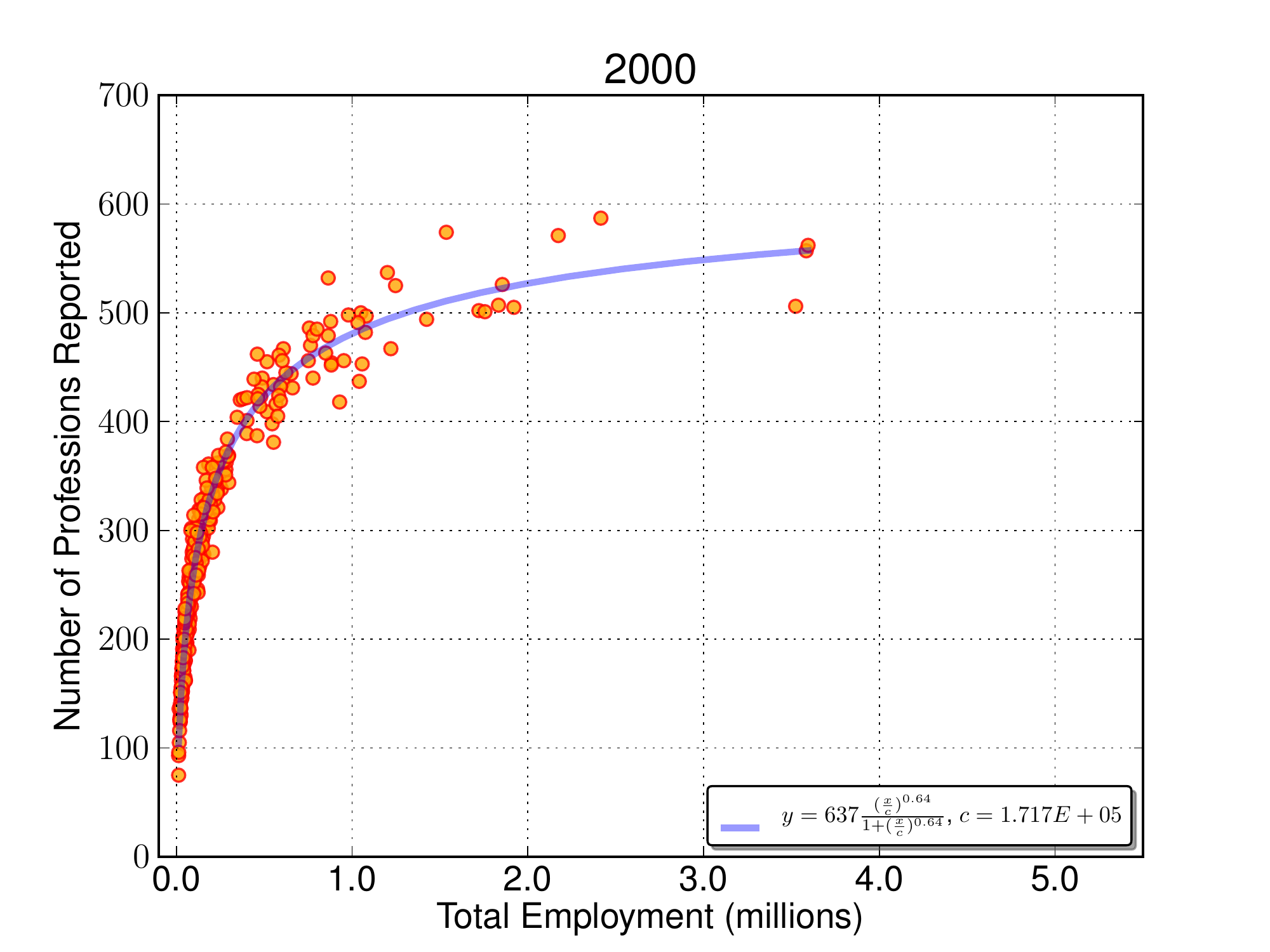}
    \includegraphics[width=0.33\textwidth]{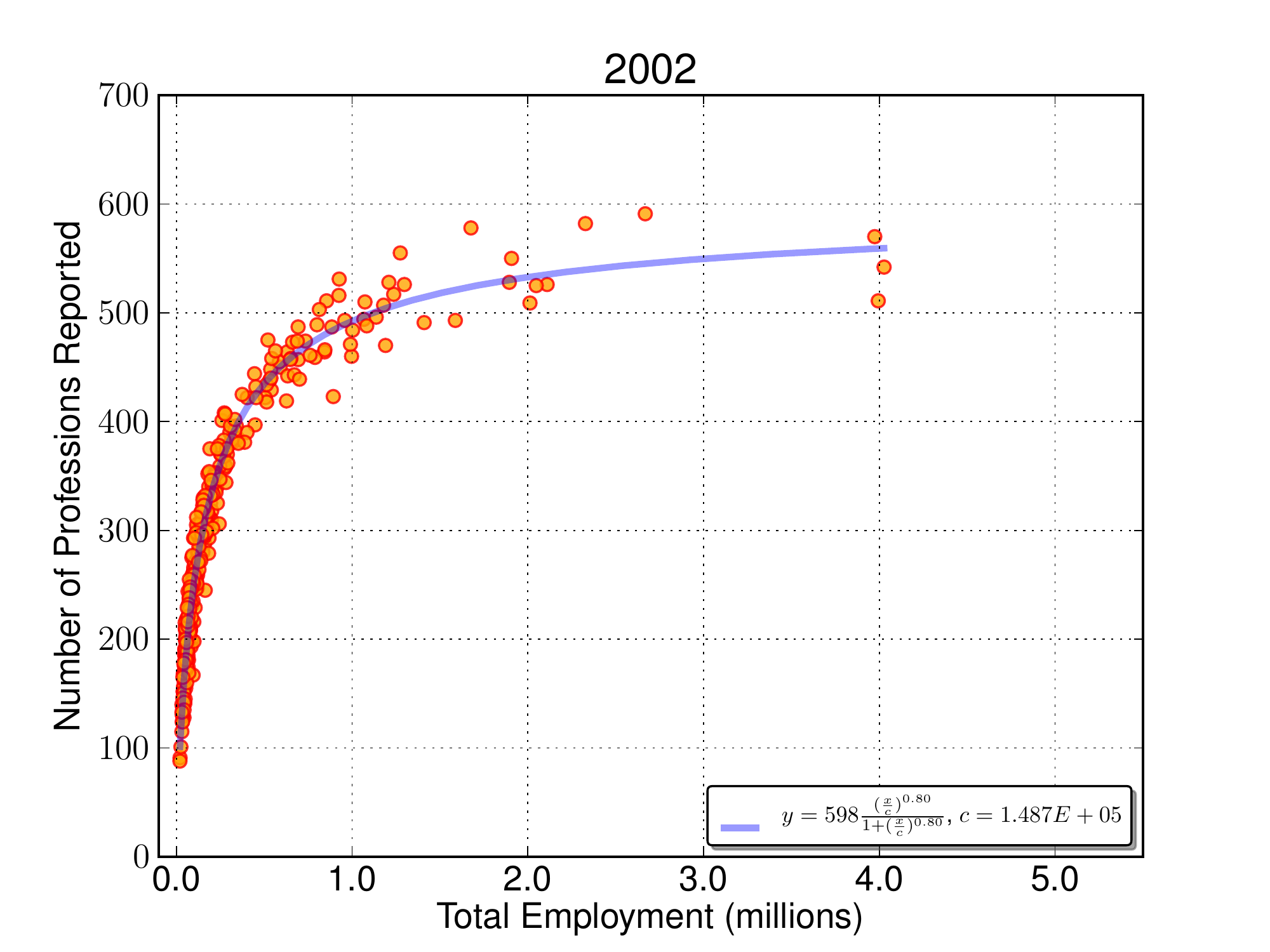}
    \includegraphics[width=0.33\textwidth]{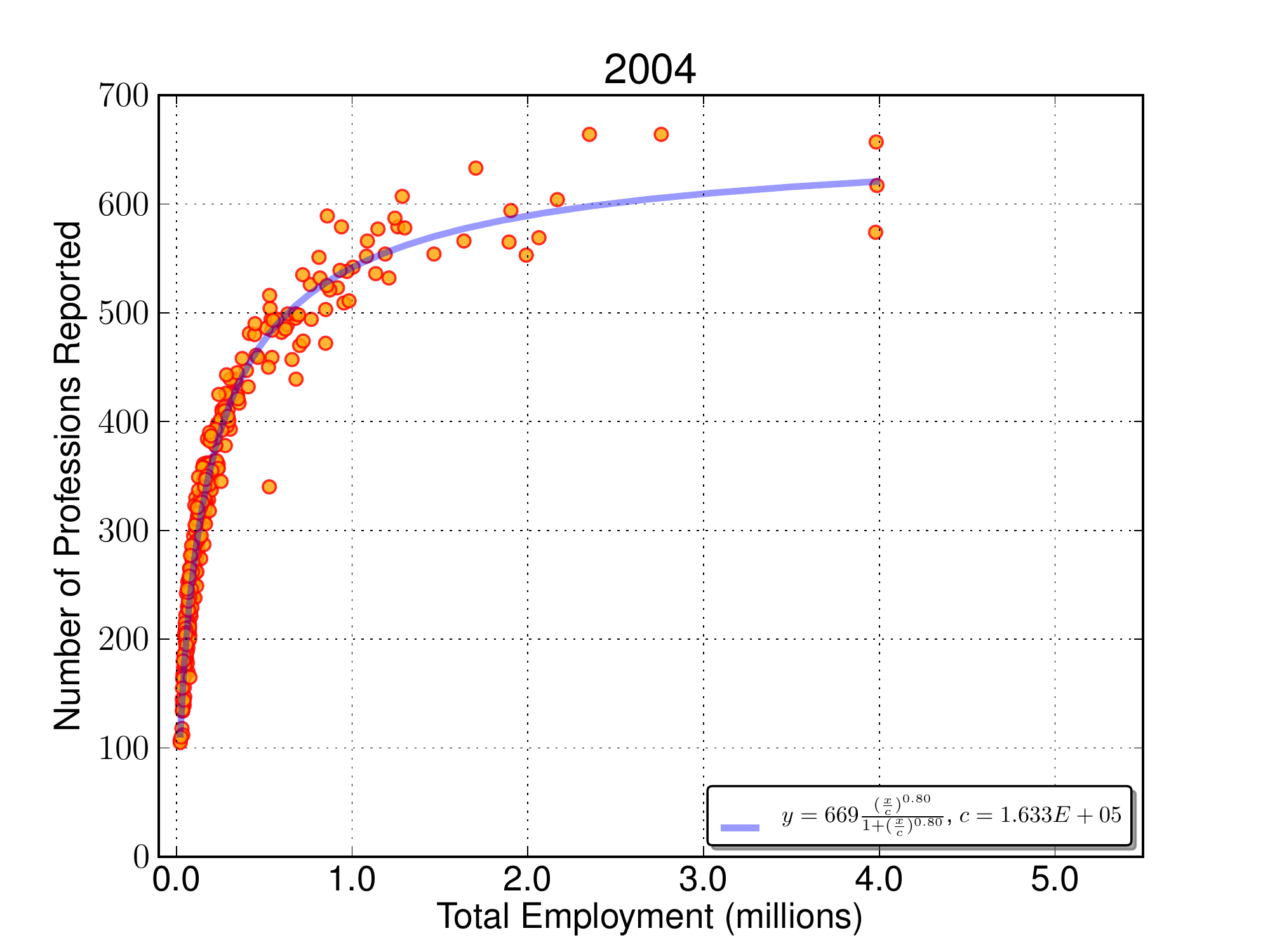}
  }
  \mbox{
    \includegraphics[width=0.33\textwidth]{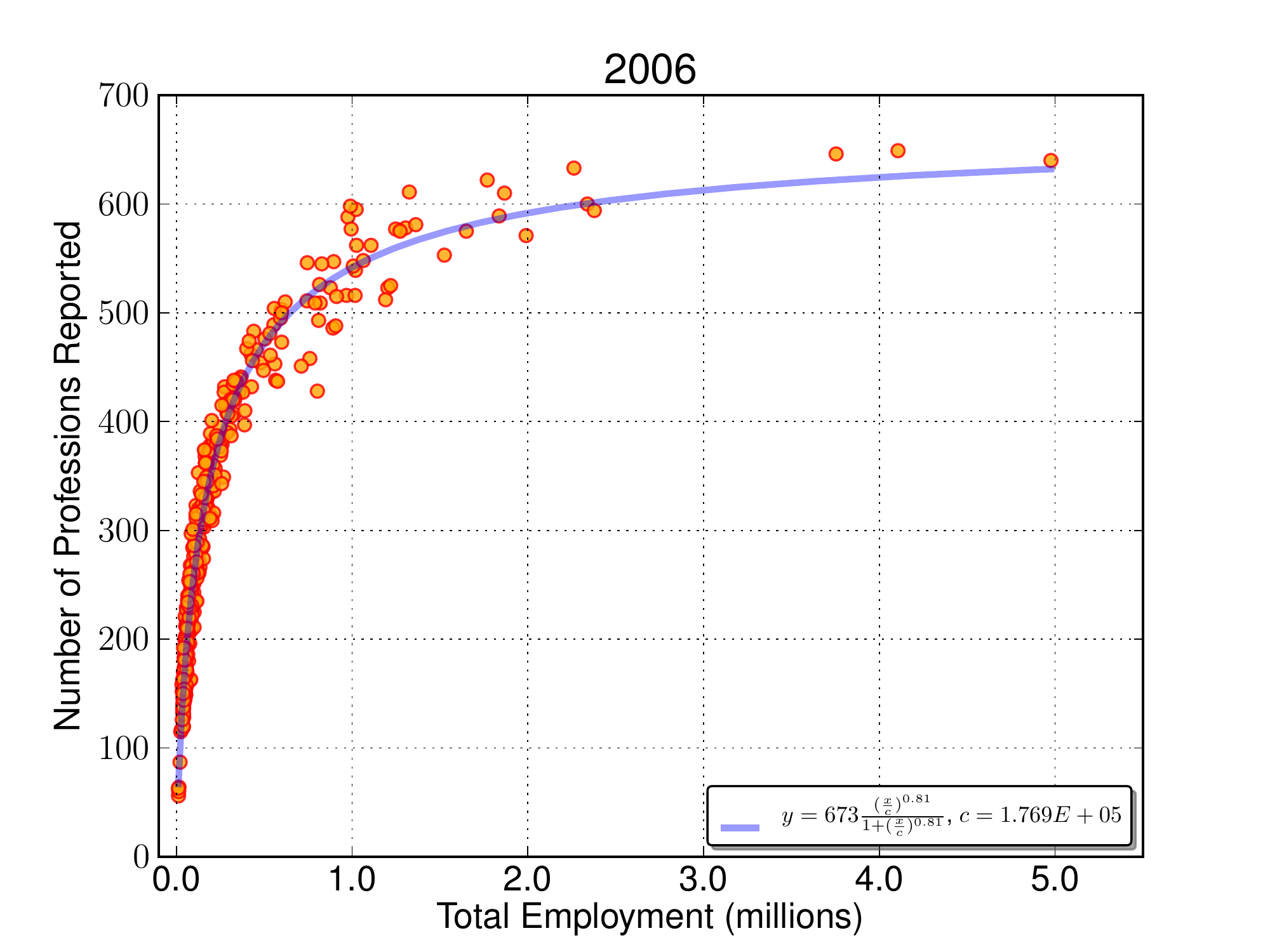}
    \includegraphics[width=0.33\textwidth]{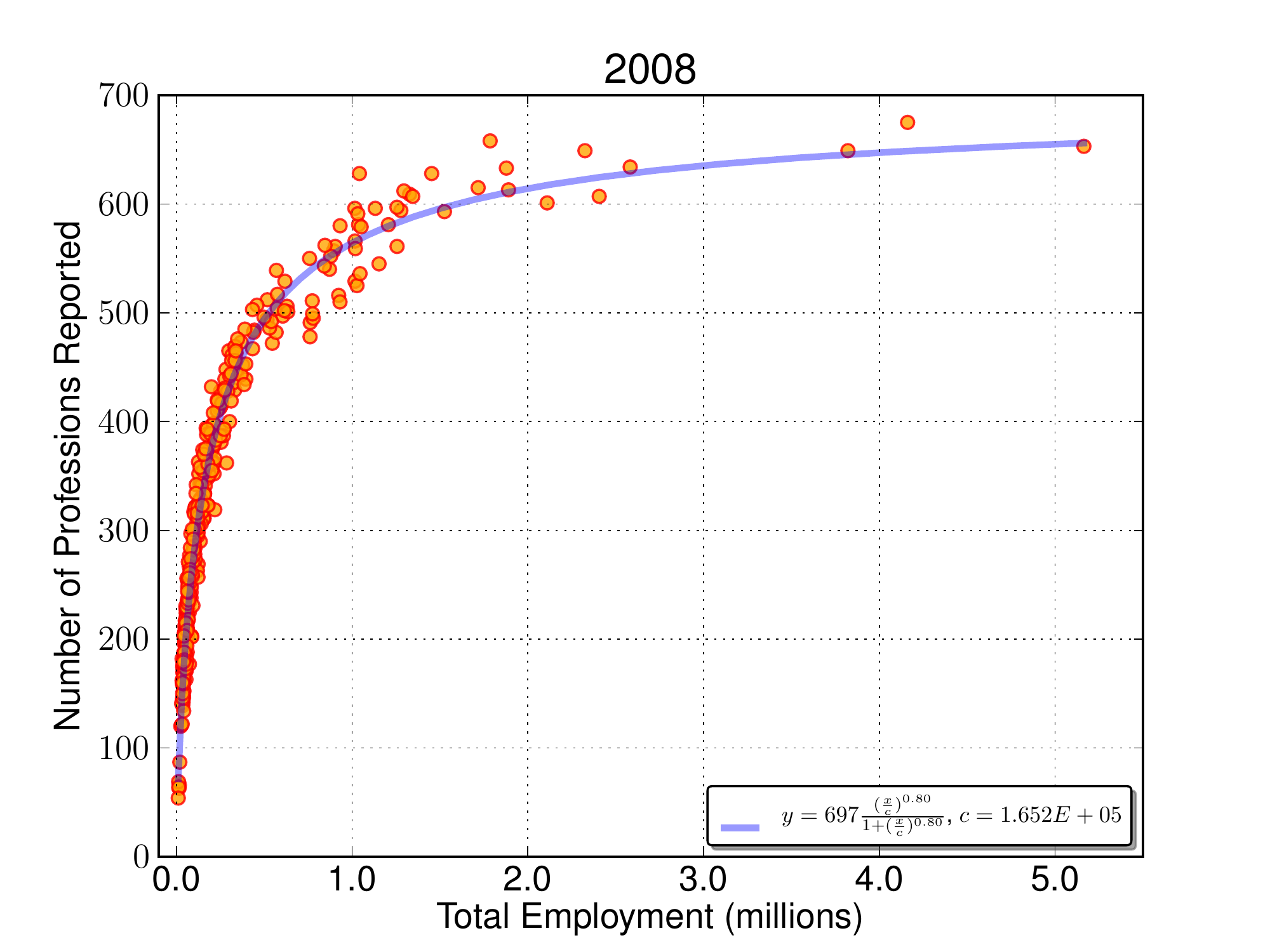}
    \includegraphics[width=0.33\textwidth]{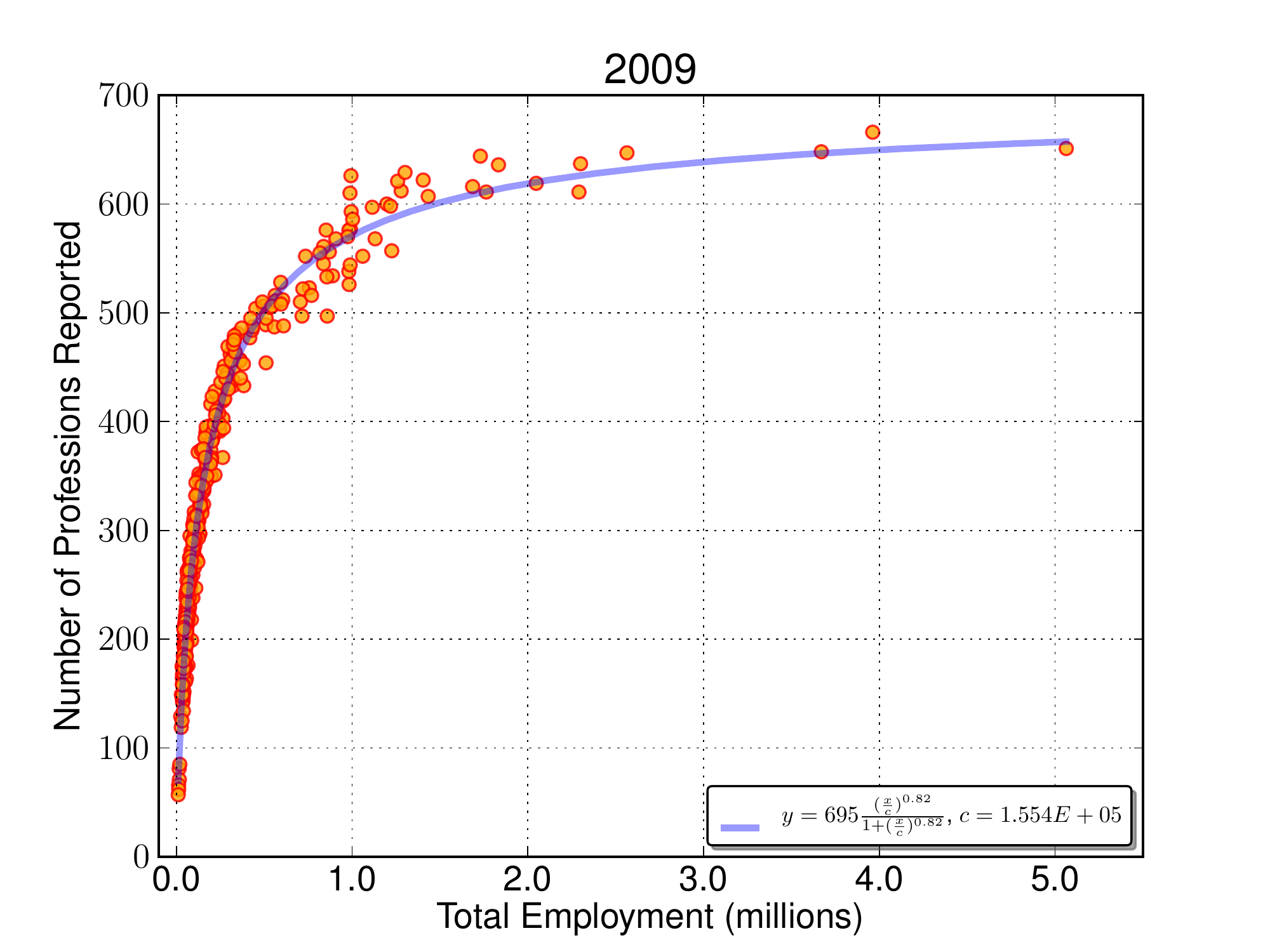}
  }
  \caption{The functional relationship between diversity of professions, $d$, and city size holds over time}
\end{figure}

\bigskip

\subsection*{Variation of $\gamma$ over time}

Variation of $\gamma$ over time is fairly stable. The last employment classification scheme was defined in 2002,
which may explain larger fluctuations (and lower values) of $\gamma$ prior to such date.

\begin{figure}[H]
  \centering
  \includegraphics[width=0.65\textwidth]{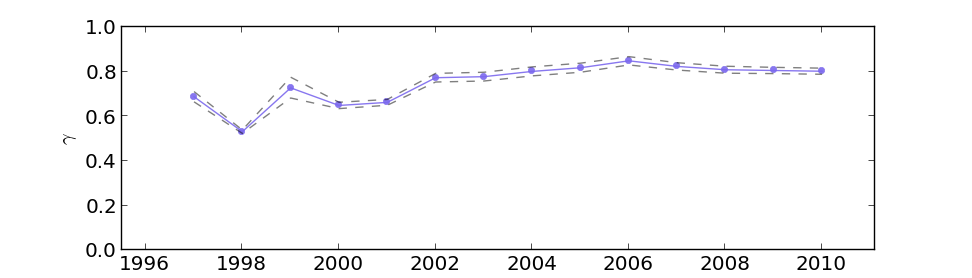}
\caption[S]{Changes in the best fit value of the exponent $\gamma$ across time. }
\end{figure}

\subsection*{Indices of diversity}

Diversity is most commonly measured in terms of functionals of the probability distribution of types, $p(i)$.
Examples of such functions are the Herfindahl--Hirschman Index ($HH$), the Shannon entropy ($S$).

The Herfindahl-Hirschman index (HH) measures how concentrated a distribution is.
For this reason it is often applied to economic sectors to measure their concentration and the creation of monopolies. Given the asymptotic form of the
distribution (Eq.~6) it can be calculated analytically as
\begin{eqnarray}
HH = \sum_{i=1}^{D(N)} p^2(i)  &=& \frac{\delta^2}{1-\delta^2} \frac{1- D_0^{-\frac{1+\delta}{1-\delta}} N^{-1-\delta}}{(1-D_0^{\frac{\delta}{1-\delta}} N^{-\delta})^2}  \nonumber \\
& \simeq & \frac{\delta^2}{1-\delta^2}  \left(1 + \frac{2}{D_0^{\frac{\delta}{1-\delta}} N^{\delta}} \right).
\end{eqnarray}
Consequently the HH index decreases towards a small constant , set by the exponent $\delta$, as cities grow. This expresses an increase in diversity 
with city size.  Note that the asymptotic value for $N \rightarrow \infty$, with $\delta
\simeq 1/6$, is $HH \rightarrow 0.028$, which is typical of highly diverse (and competitive) markets.

Similarly the Shannon entropy $S$ measures the diversity of the occupational distribution as
\begin{eqnarray}
S = -\sum_{i=1}^{D(N)} p(i) \ln p(i) \simeq \frac{1}{\delta} - D_0^{- \delta/(1-\delta)} N^{-\delta} \ln ( D_0^{1/\gamma} N ).  
\end{eqnarray}
which increases with $N$ towards the Pareto distribution limit at infinite $N$. Thus, the increase in entropy signals
the increase in diversity of the occupational distribution as cities grow.  Note that in both cases the increases in
diversity are driven, at leading order, by a term of order $N^\delta$.

\begin{figure}
  \centering
  % \begin{subfigure}{.3\textwidth}
    \includegraphics[width=.3\textwidth]{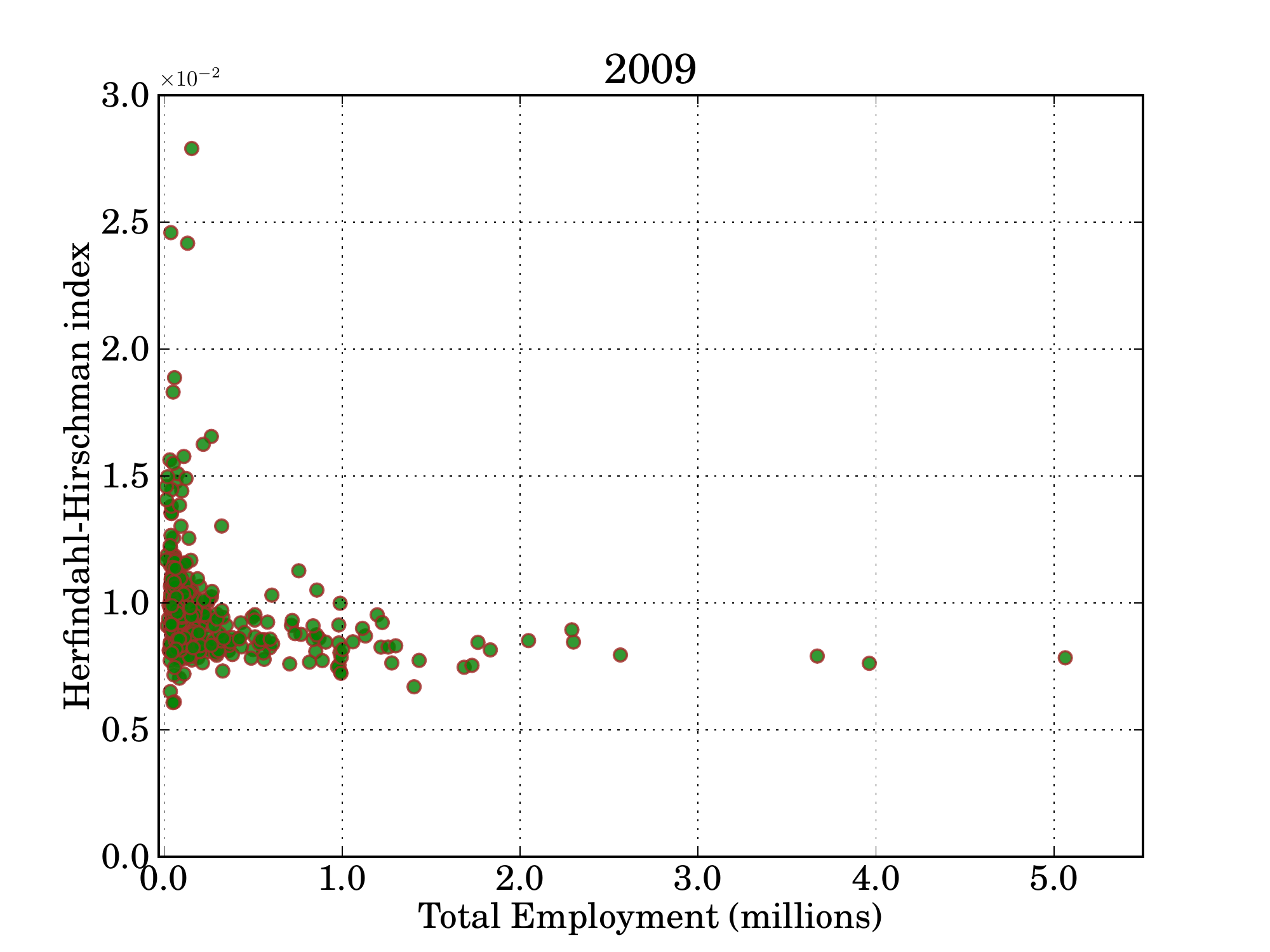}
  %   % \caption{}
  % \end{subfigure}
  % ~
  % \begin{subfigure}{.3\textwidth}
    \includegraphics[width=0.3\textwidth]{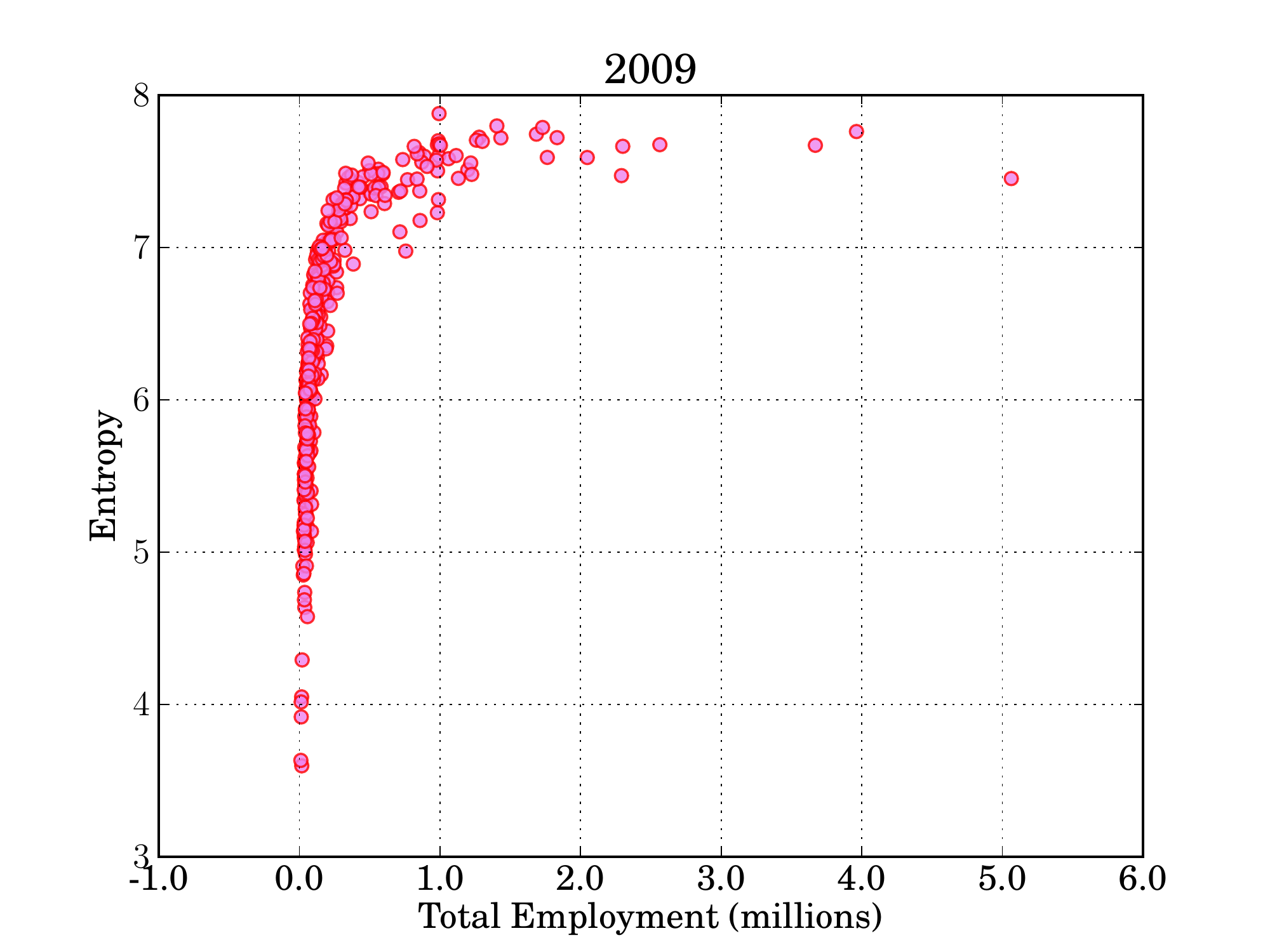}
    % \caption{}
  % \end{subfigure}
  \caption {Measures of occupation diversity in U.S. urban systems. (a) Herfindahl--Hirschman Index; (b) Shannon entropy. Both plots show the results of evaluating these indices at $r=r_6$, thus showing saturation in these measures of diversity for large cities.}
% \caption {(a) Herfindahl--Hirschman Index ; (b) Shannon entropy and (c) Kullback--Leibler divergence}
\end{figure}

\subsection*{Larger cities are more diverse, but  less diverse per capita} 

\noindent
\textbf{Theorem}\\
\noindent The number of distinct professions scales sublinearly with city population size: 

\noindent For any not fully specialized city
with $N$ inhabitants (or workers), if professions are a property of individuals that cannot be accumulated and the number of distinct professions $d$ is a scale invariant function of $N$ then its exponent $\gamma < 1$ (sublinear scaling).  \\
\textbf{Proof:} \\
We take a profession as a property of a person that cannot be accumulated. As such, the total number of distinct professions in a city $d$,
  \begin{equation}
    \frac{d}{N} \leq 1.
  \end{equation}
  From this it follows that 
  \begin{equation}
    \ln d \leq \ln N \rightarrow \frac{\Delta d}{d} \leq \frac{\Delta N}{N},
  \end{equation}
  where $\Delta$ denotes a variation. Because $d$ is a scale invariant function of $N$, $d=D_0 N^\gamma$, then
  \begin{eqnarray}
    \frac{\Delta d}{d} = \gamma \frac{\Delta N}{N},
  \end{eqnarray}
  for some real number $\gamma$, independent of $N$. Then, from (S.6) and (S.7), it follows that $\gamma \leq 1$. Because no
  city is fully specialized, the inequality holds strictly and $\gamma < 1$, as stated.

\subsection*{Productivity and mean annual wages}

We evaluated the superlinear relationship between city size and productivity, measured as the mean annual wages of each
occupation in the dataset provided by the BLS. Although these data miss a few rare professions because of reporting cutoffs due to confidentiality issues we find a super linear scaling relation with exponent $\beta=1+\delta$, $\delta = 0.18\pm 0.03$, in agreement with theoretical expectations of $\delta \simeq 1/6$ \cite{Bettencourt2012}.

\begin{figure}
  \centering
  \includegraphics[width=.7\textwidth]{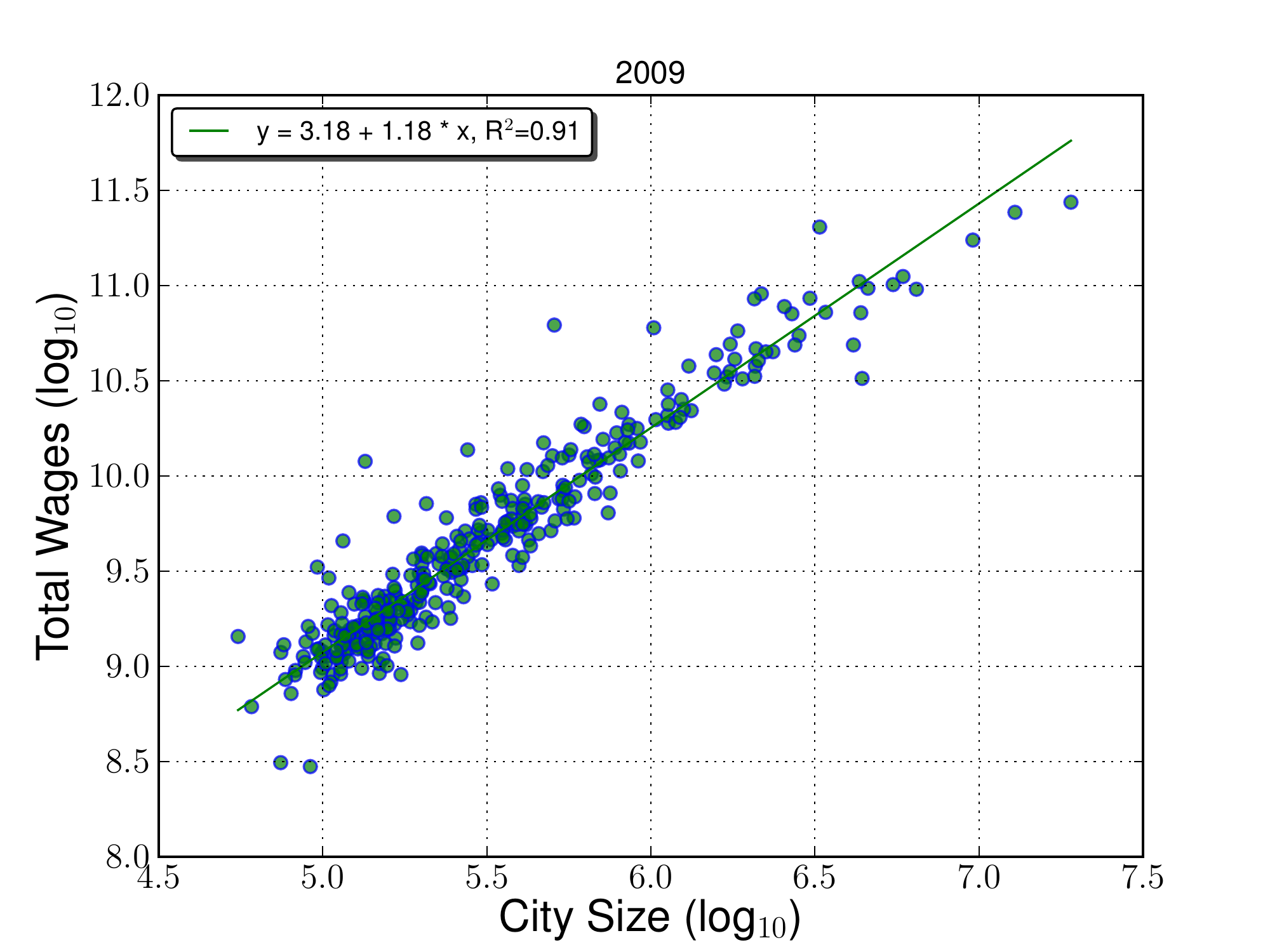}
  \caption{Total wages in US MSAs in 2009 scale superlinearly with city population size, with an exponent $\beta=1.18\pm 0.03$.}
  \label{fig:wages}
\end{figure}

\subsection*{Optimal Professional Diversity}

Here we show in greater detail how the observed open ended increase in labor productivity with city size, see previous section and Ref.~\cite{Bettencourt2012}, is related to the characterization of the changes in the levels of professional diversity reported in the main text. 

The main idea is that specialization as the source of increases in economic productivity must be accompanied by coordination of specialized tasks in order to maintain the original functionality. Thus, if an individual's tasks become more specialized, then the complementary functions shed in the process of specialization must be preserved within its socio-economic network, so that the overall function now exists in a social organization and less within the individual.  We assume that only close coordination (at least in the initial stages of transfer of functions) will be able to preserve these specialized functions suitably integrated and as such that there must be a conservation of the number of functions within the immediate contacts (first neighbors) of each individual. Thus on the average we write this condition as $k \frac{D(N)}{N} \equiv k d = A$, where $A$ is a constant in $N$, but that may vary over time, e.g. due to changes in communication and transportation technologies.   This conservation may also be able to maintain the ability of each individual to innovate through processes of functional combinations that shift functions from his own immediate responsibility to those of others with whom he is closely connected. 

The other main difference to Ref~\cite{Bettencourt2012}, is to start from the standard assumption that economic productivity is proportional to (labor) specialization, that is $w \sim 1/d$, and not necessarily to social connectivity. To preserve dimensions we need to write instead that $w = g(k d)/d$, where $g$ is a function that transforms the index of specialization $1/d$ into economic units (money per person and unit time). Its dependence on $k d$ is necessary for consistency with observations and may give interesting insights on mechanisms of economic growth over time, which however we will not address here.  

Thus, we can formulate the problem of determining the optimal professional diversity of a city with size $N$ (and consequently average per capita connectivity\cite{Bettencourt2012,Schlapfer2012} $k=k_0N^\delta$) in terms of maximizing it economic productivity subject to the constraint that activities lost to an individual remain available through its neighbors in their social network.  This can be written in terms of a standard Lagrange multiplier optimization problem,
\begin{eqnarray}
{\cal L}(d;\lambda) = \frac{ g(k d)}{d}  - \lambda \left( kd - A \right).
\end{eqnarray}
This optimization is a particular case of the problem considered in
Ref.~\cite{Bettencourt2012}, where it was shown that the socioeconomic
outputs of cities (including measures of economic output, such as
wages, GDP, etc) are proportional, on the average, to their social
connectivity, $k$. From this process we take $k(N)$ as given and show
how social connectivity and professional diversity are related.  The
solution of this optimization problem, obtained by taking the
variations of $\cal L$ relative to each of the variables, $d$ and
$\lambda$, to zero is

\begin{eqnarray}
d = \frac{A}{k} = \frac{A}{k_0} \frac{1}{N^{\delta}}, \qquad  w = \frac{g(A)}{A} k = \frac{g(A)}{A} k_0 N^{\delta}. \\
\frac{d g}{d A} - \frac{1}{A} g - \lambda_1 A = 0 \rightarrow g(A) = \left[ C + \int^A dA' \lambda_1(A') \right] A,  
\end{eqnarray}  
where $C$ is a constant of integration; $\lambda_1 = \lambda/k$ and is
assumed to be a function of $A$. Both are to be set by boundary
conditions on $g(A)$. Identifying this solution with the empirical
relations derived in the main text leads to $D_0 =
\frac{A}{k_0}$. This shows that maximizing productivity subject to the
maintenance of functionality among immediate network neighbors leads
to a prediction for the professional diversity of cities in agreement
with the findings of the main text, and shows explicitly how scale,
productivity, functional diversity and network structure must be
integrated in understanding the evolution of urban economies with
population size.

\bigskip

\bibliographystyle{Science}
 
% \bibliography{Urban_Dynamics} 